\documentclass[11pt]{article}

\usepackage{graphicx}

\usepackage{multicol} 

\usepackage{color}
\usepackage{latexsym}

\usepackage[utf8]{inputenc}

\definecolor{rosso}{cmyk}{0,1,1,0.4}
\definecolor{rossos}{cmyk}{0,1,1,0.55}
\definecolor{rossoc}{cmyk}{0,0.5,1,0.2}
\definecolor{blu}{cmyk}{1,1,0,0.3}
\definecolor{blus}{cmyk}{1,1,0,0.6}
\definecolor{blucc}{cmyk}{1,0.4,0.2,0}
\definecolor{viola}{cmyk}{0,1,0,0.6}
\definecolor{viola2}{cmyk}{0,1,0.2,0.6}
\definecolor{verde}{cmyk}{0.92,0,0.59,0.25}
\definecolor{verdec}{cmyk}{0.92,0,0.59,0.15}
\definecolor{verdes}{cmyk}{0.92,0,0.59,0.4}
\font\tenrsfs=rsfs10 at 12pt
\font\sevenrsfs=rsfs7
\font\fiversfs=rsfs5
\newfam\rsfsfam

\textfont\rsfsfam=\tenrsfs
\scriptfont\rsfsfam=\sevenrsfs
\scriptscriptfont\rsfsfam=\fiversfs
\def\mathscr#1{{\fam\rsfsfam\relax#1}}

\def\circa#1{\,\raise.3ex\hbox{$#1$\kern-.75em\lower1ex\hbox{$\sim$}}\,}

\usepackage{amsmath,latexsym,amssymb,color,hyperref,graphicx}
\newcommand{\eq}[1]{(\ref{#1})}
\newcommand{\be}{\begin{equation}}
\newcommand{\ee}{\end{equation}}
\newcommand{\bea}{\begin{eqnarray}}
\newcommand{\ena}{\end{eqnarray}}
\newcommand{\no}{\noindent}
\newcommand{\nb}{\nonumber}

\renewcommand\l{\lambda}
\renewcommand\o{\omega}

\renewcommand\l{\ensuremath{\lambda}}

\newcommand\m{\ensuremath{\mu}}

\newcommand\n{\ensuremath{\nu}}

\newcommand{\de}{\partial}
\newcommand{\ha}{\frac{1}{2}}

\renewcommand\l{\ensuremath{\lambda}}

\newcommand{\ba}{\begin{eqnarray}}
\newcommand{\ea}{\end{eqnarray}}
\newcommand{\plm}{M_{\text{Pl}}} 
 
\makeatletter
\def\ps@mine{%
    \def\@oddfoot{\hfil\thepage\hfil}\let\@evenfoot\@oddfoot
    \let\@oddhead\@evenhead%
    \let\@mkboth\@gobbletwo
    \let\sectionmark\@gobble
    \let\subsectionmark\@gobble
    }
\renewcommand\section{\@startsection {section}{1}{\z@}%
                                   {-3.5ex \@plus -1ex \@minus -.2ex}%
                                   {2ex \@plus.2ex}%
                                   {\normalfont\large\sffamily\bfseries}}
\renewcommand\subsection{\@startsection {subsection}{1}{\z@}%
                                   {-3.5ex \@plus -1ex \@minus -.2ex}%
                                   {2ex \@plus.2ex}%
                                   {\normalfont\sffamily\bfseries}}
\makeatother

\numberwithin{equation}{section}

\begin{document}
\thispagestyle{empty}
\vspace*{-2.5cm}
\begin{minipage}{.45\linewidth}
\begin{flushleft}                           
%{\footnotesize SACLAY-t15/xxx\\CERN-PH-TH-2015-xxx}
\end{flushleft} 
\end{minipage}
\vspace{2.5cm}

\begin{center}
{\huge\sffamily\bfseries 
On the 6th Mode in Massive Gravity
 }
 \end{center}
 
 \vspace{0.5cm}
 
 \begin{center} 
 {\sffamily\bfseries \large  Marco Celoria }$^a$  {\sffamily\bfseries  \large Denis Comelli}$^b$, {\sffamily\bfseries \large Luigi Pilo$^{c,d}$}\\[2ex]  {\it
$^a$ Gran Sasso Science Institute (INFN)\\Via Francesco Crispi 7,
L'Aquila, I-67100\\\vspace{0.1cm}
$^b$INFN, Sezione di Ferrara,  I-44122 Ferrara, Italy\\\vspace{0.1cm}
$^c$Dipartimento di Scienze Fisiche e Chimiche, Universit\`a dell'Aquila,  I-67010 L'Aquila, Italy\\\vspace{0.1cm}
$^d$INFN, Laboratori Nazionali del Gran Sasso, I-67010 Assergi, Italy\\\vspace{0.3cm}
{\tt marco.celoria@gssi.infn.it}, 
{\tt comelli@fe.infn.it}, {\tt luigi.pilo@aquila.infn.it}
}
\end{center}
   
\vspace{0.7cm}

\begin{center}
{\small \today}
\end{center}

\vspace{0.7cm}

\begin{center}
{\sc Abstract}

\end{center}
\no Generic massive gravity models in the unitary gauge correspond  to
a self-gravitating medium with six degrees of freedom.  It is widely
believed that massive gravity models with  six degrees of freedom have an unavoidable ghost-like
instability;  however, the corresponding medium has stable phonon-like
excitations. The apparent contradiction is
solved by the presence of a non-vanishing background pressure and
energy density of the medium that opens up a stability window. The
result is confirmed by looking at linear stability on an expanding
Universe, recovering  the flat space stability conditions in the
small wavelength limit. Moreover, one can show that under
rather mild conditions, no ghost-like instability is present for any
wavelength. As a result, exploiting the medium interpretation, a generic
massive gravity model with six  degrees of freedom is perfectly viable.

  \vspace{2.cm}
\section{Introduction}
In recent years there has been a renewed interest in massive gravity,
see for instance the
reviews~\cite{Rubakov:2008nh,Hinterbichler:2011tt,deRham:2014zqa}.
One the main issue of massive gravity is that a randomly picked model
propagates six degrees of freedom (DoF) and among the two scalar modes
one leads to ghost instabilities~\cite{deRham:2014zqa}. A lot of
effort has been devoted in finding suitable models in which only five
healthy DoF are present. 
At the non-perturbative level, when a residual
Lorentz symmetry is imposed in the massive gravity action, only the
dRGT model~\cite{deRham:2010kj} has such a feature; if only only
rotational symmetry is imposed,many choices are available~\cite{Rubakov:2004eb,Comelli:2013txa,Comelli:2014xga}. 
Unfortunately in both cases the special choice of the ghost free action is not protected
by symmetries. In the Lorentz invariant case there is a number of additional
phenomenological difficulties related with the absence of a spatially
flat Friedmann-Lemaitre-Robertson-Walker (FLRW) cosmological solutions,
the presence of the vDVZ
discontinuity~\cite{vanDam:1970vg,Zakharov:1970cc,Vainshtein:1972sx} and serious
difficulties with unitarity~\cite{Bellazzini:2017fep}.
In general, the recent detection of gravitational
waves~\cite{TheLIGOScientific:2017qsa,Monitor:2017mdv} poses strong
constraints on modified gravity theories relevant for dark energy, see for
instance~\cite{Creminelli:2017sry,Sakstein:2017xjx}, disfavouring
theories in which gravitational propagation velocity is modified.

In this paper we reconsider the faith of the sixth mode in the light of
the interpretation of massive gravity as a self-gravitating
medium~\cite{Dubovsky:2005xd,ussgf,classus}. 
The crucial point is that
the energy density and pressure of the medium when present can open up
a region of stability for massive gravity models with healthy six DoF. The very
same mechanism guaranties that that two scalar phonon-like modes of perfect
fluid are stable; the argument can be extended to a general medium.
\\
Technically we follow the following path
\begin{itemize}
\item A generic massive gravity models with six DoF  around
Minkowski is intrinsically plagued by a scalar ghost.
\item
In the medium interpretation of the massive gravity models we 
find that
in the absence of dynamical gravity, the phonon like excitations of such a models are stable.
\item
 For consistency, the presence of the medium  requires a
non zero energy density  $\rho$ and pressure  $p$,
  which, when gravity is switched on, leads to the departure from flat
  space as a consistent background.
\item Analysing perturbations around a FLRW background
in the short wave length limit, the same stability
conditions for a medium on Minkowski spacetime are recovered.
\end{itemize}
This shows that massive gravity models with six DoF can be intrinsically stable. 
 
 The outline of the paper is the following. In section
   \ref{MMS} we study the stability of a generic medium in Minkowski
   space. In section \ref{self} we relate massive gravity with a
   self-gravitating medium b exploiting the unitary gauge. Section
   \ref{FLRW} is devoted to the analysis of stability  for a self-gravitating medium
   around a FLRW background. Our conclusions are given in section \ref{conc}.

\section{A Medium in Minkowski Space}
\label{MMS}
Let us   start  from a scalar field theory of the form 
 \be
 S_m= \int d^4x\sqrt{-g} \; U(\de \varphi)\,;
\label{fact} 
\ee
where  $U$ is
the Lagrangian for four derivatively coupled scalar fields
$\varphi^A$, $A=0,1,2,3$  invariant under shift symmetry
$\varphi^A\rightarrow \varphi^A+c^A$ with $ \quad \partial_\mu
c^A=0$. As discussed in detail
in~\cite{Dubovsky:2005xd,Dubovsky:2011sj,Ballesteros:2012kv,ussgf,classus}
such an action can be used to give an effective description of the
dynamics of self-gravitating media coupled with
gravity. The low energy excitations (phonons) can be described by
suitable fluctuations of the $\varphi^A$, 
see appendix \ref{medium} for details. 
In this chapter we  consider the
perturbations $\pi^A$ around flat space 
\be
\varphi^A = \bar{\varphi}^A + \pi^A \, .
\ee
It is easy to see that in order to have a EMT  constant, the background value of the scalar fields is $
\bar{\varphi}^A =x^A$. The presence of the fluid breaks spontaneously the
Lorentz group.
Notice that setting $\bar \varphi^A=0$ it would imply that there is no background
fluid ($\rho=p=0$).
 The  Lagrangian  $U$  gives the following  background EMT 
\be\label{EMT}
 T^{\mu\nu}_{\text{Mink}} = %\rho\, u^\mu \, u^\nu + p \, h^{\mu \nu} =
  (\rho \;\delta^\mu_0\;\delta^\nu_0\;  +
 p\;\delta^\mu_i\;\delta_{ij}\;\delta^\nu_j ) =p\;\eta^{\m\n}+(\rho+p)\;u^\mu\;u^\nu\, ;
  \ee
where $\rho$ and $p$ are constant that gives the energy density and the pressure  evaluated on flat space by \eq{prho}, measured by observers with 4-velocity $u^\mu=(1,\vec{0} \, )$. \\
The first thing to be noticed is that if one requires   the background EMT to be Lorentz
 invariant~\footnote{Of course this is statement based on symmetry and
not on the equations of motion.} 
then we have  to impose $\rho+p=0$, i.e. $T^{\mu\nu}_{\text{Mink}}=p\;\eta^{\m\n}$, i.e. the fluid
has the same equation of state of a cosmological
constant. 
From the media/fluid interpretation it is
rather natural that the {\it vacuum} state breaks Lorentz by its
presence, that is precisely what happen in any solid state lab. In
what follows we will consider the case $\rho+p \neq 0$, while the
exotic case $\rho+p = 0$ will be studied elsewhere.\\
Decomposing the $\pi^A$ fields according to  $SO(3)$
vector and scalar perturbations:    $\varphi^0=t+\pi_0$  and
$\varphi^i=x^i+\partial_i \,\pi_l+V^i$ (with $\partial_i\,V^i=0$),
 we get that for the scalar modes the quadratic Lagrangian is\footnote{We can get
   such a Lagrangian from equations \eq{KN}   taking only the 3rd and 4th
   diagonal entries and then imposing the Minkowski limit  ($\phi'=a=1,\;{\cal H}=0$).}
\be
L_{\pi}^{(s)} = \frac{k^2}{2} \left( \hat{M}_1 +p+ \rho\right)\pi_l'{}^2 +
 \hat{M}_0 \,  \pi_0'{}^2 + \frac{k^2}{2} \, \hat{M}_1
\pi_0^2 + k^2 \left(\hat{M}_1-2 \, \hat{M}_4\right) \pi_0' \, \pi_l 
+ k^4 \left(\hat{M}_3-\hat{M}_2\right) \pi_l^2 \, ;
\label{sact}
\ee
where we have introduced the mass paratemeters $M_i$ defined in terms
of the derivatives of the medium Lagrangian (\ref{fact}), their explicitly
expression (\ref{MMI}) is given in appendix~\ref{medium}; such parameters
are quite useful in order to classify the various media
with respect of their mechanical  and  thermodynamical properties~\footnote{From the
  mechanical point of view we have: perfect fluids with $M_{1,2}=0$,
  solids $M_1=0$, superfluids $M_2=0$ and supersolids
  $M_{1,2}\neq0$. From the thermodynamical point of view we have:
  adiabatic media when $M_1=0$, isentropic media when $M_0=0$,
 isentropic perfect fluids when $
  \hat M_1+\rho+p=0$.}. Moreover it  is convenient to set $\hat M_a =
\plm^2 \, M_j$, $j=0,1,2,3,4$. The  linear expansion of the
medium action gives
\be
S_{\pi}^{(1)}=\int \;d^4x\;\left[\sigma \;
  \dot{\pi}^0+(\sigma-p-\rho)\, \de_i\pi^i \right]=0 \, ;
\ee
where $\sigma$ is the background entropy per particle.  By integration
by parts, the linear action is zero upon the constancy in space and in
time of
$\sigma$. Thus no condition on $p$ and $\rho$ needs 
to be  imposed in sharp contrast with the case where dynamical gravity is present, as
we shall see later.

For transverse vectors  $\vec{V}=(V^1,V^2,V^3)$ one gets
\be
L_\pi^{(V)} =\frac{1}{2} \left(\hat{M}_1+p+ \rho
\right) \;\vec{V}'^2 -  k^2\;
\hat{M}_2 \; \vec{V}^2 \, .
\label{vact}
\ee
We consider the case of generic $M_i$ values, thus no  tuning and/or
additional symmetries are  imposed. From the kinetic terms in
\eq{sact}  it is clear that there are two critical points $\hat M_1+p+\rho=0$ and $\hat M_0=0$ where at least one DoF can be integrated  and the stability analysis has to be redone.
 As soon as we are away from such a points we can study the
stability looking to the total energy, or equivalently the Hamiltonian
(see appendix \ref{stabcomp}). In the scalar sector the
  total energy is  given by
\be
E_s= \hat{M}_0\; \pi _0'{}^2+\frac{k^2}{2} \, 
   \left(\hat{M}_1+p+ \rho\right) \pi _L'{}^2+ k^4 \, 
   \left(\hat{M}_2-\hat{M}_3\right) \pi _L^2- \frac{k^2}{2}
   \hat{M}_1 \,  \pi _0^2 \, .
\label{energy}
\ee
Imposing that the the energy  is  bounded from below in both the scalar and
vector sectors leads to (see appendix \ref{stabcomp})
\be
\begin{split}
& M_0 >  0  \, , \quad
-(p+ \rho) < \hat{M}_1 < 0  \, , \qquad M_2 > 0  \, , \qquad
M_2 > M_3 \, .
\end{split}
\label{stabc}
\ee
Clearly, when $\rho+p >0$,  stability is possible,  as it
should be. Actually the conditions of dynamical and
thermodynamical stability for a perfect fluid are the
same~\cite{usthermo}.   It is interesting to note that the previous
stability conditions do not hold in the Lorentz invariant
case~\footnote{Here we do not consider that the case where
  $\bar{T}^{\mu \nu}$ is not Lorentz Invariant but the quadratic
  action Lagrangian is Lorentz invariant.}. Indeed,
requiring that that $T_{Mink}^{\mu \nu}$ to be proportional to
$\eta^{\mu \nu}$ implies that $p+\rho=0$; thus only the barotropic
equation of state $w=-1$ is viable and the stability windows
closes down~\footnote{Unless we require $M_0=0$ or $\hat{M}_1 +\rho
  +p=0$ resulting in less then six propagating DoF.}.  
  
Though St\"uckelberg fields are perfectly suitable for studying
stability,  they are not invariant under the shift symmetries
of the action and their physical interpretation is  not direct. It is convenient to introduce~\cite{usthermo,classus} as observables: the energy density perturbation $\delta \rho$, the
pressure perturbation $\delta p$  and the entropy per
particle perturbation $\delta \sigma$ given by
\be
\begin{split}
&\delta \sigma = 2 \; \left(\hat{M}_0\;\pi_0{}' -
   \hat{M}_4\; k^2 \, \pi_l \right) \, , \qquad \delta
\rho = \delta \sigma - (p+\rho) k^2 \, \pi_l , \\
& \delta p = c_s^2 \, \delta \rho + (c_b^2 -c_s^2) \, \delta \sigma \, ;
\end{split}
\ee
where
\be
 c_b^2= - \frac{M_0}{M_4} \,,  \qquad c_s^2= \frac{6 \, \hat{M}_4^2+2 \,
  \hat{M}_0 \left( \hat{M}_2-3 \,  \hat{M}_3\right)}{3 \, \hat{M}_0 \, 
  (p+\rho )} \, .
\ee
The equations of motion derived from (\ref{sact}) can be written as
\be
\delta   \sigma '' + c_\sigma^2 \,  k^2 \, \delta \sigma+\frac{
  \left(c_\rho^2- c_b^2  \right) \hat{M}_1}{\left(\hat{M}_1+p+\rho
   \right)}  \, k^2 \, \delta \rho   =0 \, ;
\label{eqsigma}
\ee
\be
\delta \rho '' + k^2 \, c_\rho^2 \, \delta \rho + k^2 \,  \left(c_b^2-c_\rho^2\right) \delta \sigma=0 \, ;
  \ee
with
\be
c_\sigma^2 =-\frac{\hat{M}_1 \left[\left(2 \, \hat{M}_4+p+\rho
    \right){}^2+4 \, \hat{M}_0
   \left(\hat{M}_2-\hat{M}_3\right)\right]}{2 \, \hat{M}_0 \, (p+\rho ) \left(\hat{M}_1+p+\rho
   \right)} \, , \quad c_\rho^2 = c_s^2+\frac{4  \, \hat{M}_2}{3 \, (p+\rho
   )} = 2\; \frac{\hat{M}_4^2+\hat{M}_0
     \left(\hat{M}_2-\hat{M}_3\right)}{\hat{M}_0 \, (p+\rho)} \, .
\ee
The evolution equation (\ref{eqsigma}) has been derived in the
hypothesis that $M_1 \neq 0$. If  $M_1= 0$ we simply have
\be
\delta \sigma' =0  \, .
\ee
It is clear from the equations of motion that the second scalar mode,
beside $\delta \rho$ is the entropy per particle $\delta \sigma$. 
The stability condition that we have found imply that the speed of sound
$c_\rho^2$ of $\delta \rho$ and $c_\sigma^2$ of $\delta \sigma$ are
both positive. Generically, the two scalar perturbations of a media are the energy
density $\delta \rho$ and the entropy per particle $\delta \sigma$ and
their dynamics is stable.

  %%%%%%%%%%%%%%%%%%%
  %%%%%%%%%%%%%%%%%%%
  
\section{Self-gravitating Medium and Massive Gravity}
\label{self}
Let us now include dynamical gravity in our picture. The medium
becomes self-gravitating by simply coupling minimally the scalar
fields with gravity. Namely, one consider the following action 
\be
 S=\plm^2\;\int d^4 x\;\sqrt{-g}\, R+\int d^4x\sqrt{-g} \; U(\de \varphi,\;g_{\m\n})\,;
\label{factg} 
\ee
where $R$ is the Ricci scalar, 
% mass;
as before,  $U$,  is the medium Lagrangian containing  four
derivatively coupled scalar fields minimally coupled with gravity (see
appendix~\ref{medium} for more details). Perturbations around flat
space are studied by taking 
\be
ds^2 = \eta_{\mu \nu} + h_{\mu \nu} \, dx^\mu
dx^\nu \, ,  \qquad\qquad
\varphi^A = \bar{\varphi}^A + \pi^A \, .
\ee
By definition,
in the unitary gauge the scalar fields
fluctuations are gauged away and their value coincides with the
background values $\varphi^A_{(U)} = x^\mu\;\delta_\mu^A$;
all perturbations  are in the  metric and we
set $h_{\mu \nu} = h_{\mu\nu}^{(U)}$.
In flat space, the expansion of the action \eq{fact} 
generates a linear term
\be\label{UEMT}
{\cal L}_1^{(U)}\equiv \frac{1}{2}\;T^{\mu \nu}_{\text{Mink}} \;
  h_{\mu \nu}^{(U)} 
\ee
proportional to the EMT of the medium $ {T}^{\mu  \nu}_{\text{Mink}}$   \eq{EMT}. 
At the quadratic level there is a contribution coming from both the
Einstein-Hilbert Lagrangian, ${\cal L}_{g}^{(U)}$ and the medium
action, the later  can be parametrised in the form 
\cite{Rubakov:2004eb,Dubovsky:2004sg,Rubakov:2008nh,Blas:2009my}
\be\label{LUm}
\left.{\cal L}_{m}^{(U)}  =\sqrt{-g}\;U\right |_{{\cal O}(2)}=
\frac{\plm^2} {4} \left[\lambda_0^2 \;  h_{00}^{(U)}{}^2+2\,\lambda_1^2\;  h_{0i}^{(U)\,2}
    -2\,\lambda_4^2 \; h_{00}^{(U)}\,  h_{ii}^{(U)} +\lambda_3^2 
\;  h_{ii}^{(U)}{}^2  - \lambda_2^2\;   h_{ij}^{(U)\,2} %\,  h_{ij}^{(ug)}
\right] ,
\ee
whose structure is the origin of the name {\it massive gravity
theories}. Thus in the unitary gauge a self-gravitating medium is
equivalent to a general massive gravity theory.
The parameters $\l_i$ are  given in terms of suitable
  first and second derivatives of the function $U$
computed on Minkowski background and their expression can be found
in~\eq{LLI} and are simply related to  the parameters $M_i$ used to
study the dynamics of a medium~\cite{ussgf,classus} by  
 \be
\label{MM}
M_0= \lambda_0^2-\frac{\rho}{2\;\plm^2},\qquad
 M_{1,2}=\lambda_{1,2 }^2 - \; \frac{ p}{\plm^2} ,
\qquad
 M_{3,4 }=\lambda_{3,4 }^2 -\frac{  p}{2\;\plm^2} \, .
 \ee
To be as general
as possible we imposed on the mass terms only rotational invariance.
Such a choice is rather natural in
the medium picture the spontaneous breaking of 
Lorentz symmetry originates simply by the presence of the medium.
Strictly speaking  only the presence of the term
$\lambda^2_2$ corresponds to  a mass term for the spin two component
of the graviton, however we will generically  refer to massive gravity
when at least one of the  $\lambda_i^2$ is non-vanishing.
Insisting on having a Lorentz symmetric background configuration
constraints the value  the $\lambda_a$ and the quadratic expansion~\ref{LUm}
reduces to
\be
{\cal L}^{(2)}_{\text{mass}}  =\frac{\plm^2} {4} \left[  A \; h_{\mu \nu }^{(U)}\;
  h_{\alpha \beta}^{(U)} \; \eta^{\mu \alpha} \; \eta^{\nu \beta} +B \;
  \left( h_{\mu \nu }^{(U)} \; \eta^{\mu\nu} \right)^2\right] \, ;
\label{masst}
\ee
thus $\lambda_0^2 = A+B$, $\lambda_1^2 =\lambda_2^2=-A$ and
$\lambda_3^2=\lambda_4^2 = B$. Of course in this case the
corresponding medium is rather special; Lorentz invariance forces to have
$p=-\rho$ and thus $T^{\mu \nu}_{\text{Mink}}= - \rho \, \eta^{\mu
  \nu}$.\\
Before proceeding let us point out that a
consistent study of gravity perturbations around flat space conflicts with the
fluid/medium picture. Take for instance the unitary gauge;
flat space is a solution of the background equations only if $\rho=p=0$,
as a result no fluid is present.\\ 
Physically it simply means that when
gravity matters, the consistent background is not flat:
the presence of a non-vanishing $p$ and $\rho$ wants to warp 
the background. To study perturbations in a consistent way, keeping $p$ and $\rho$
non-vanishing, one has to look carefully at the scale involved.
\begin{enumerate}
\item 
If the curvature radius is large compared to the typical wavelength of
the perturbations we are considering, the flat space picture is
adequate and the fluctuations of the spacetime metric can be neglected.
\item
The curvature radius is comparable with the typical wavelength of the perturbations, as a result
the background solution has to amended and the metric fluctuations are important.
\end{enumerate}
In case 1, at very large momentum and
energy, curvature is negligible together with the mixing of the
St\"uckelberg fields with the metric; much like in a spontaneously
broken gauge theory where the ultraviolet  behaviour is captured by
$\pi$s alone. Thus, one can forget about gravity and simply study the effective
quadratic Lagrangian for the $\pi$s obtained by expanding $U$
and the
stability analysis is the one given in the previous section. When case 2
applies we need to consider a consistent background, the natural one
is an expanding FLRW Universe that will be discussed in the next
section.\\

One may be tempted to simply forget about the
restriction imposed by the background equations, however this is not an
option and it leads to a contradiction. Indeed, as shown in appendix
\ref{stabcomp},  the quadratic Lagrangian  in unitary gauge does not
depend on $p$ and $\rho$ and for generic values of the masses two DoF
are present and one is a ghost. On the other hand,
employing the Newtonian gauge, the quadratic action depends on both
$p$ and $\rho$ but this time, though the kinetic matrix for the
relevant fields is not positive definite,  three DoF seems 
to be present. As expected, to  reconcile the two gauges one has
that $\rho=p=0$. The bottom line is that, a self-gravitating
medium cannot be consistently studied in flat space unless there is no
medium !  We stress again that the medium picture is not available
when Lorentz invariance is imposed on the background EMT, then
$p+\rho=0$. In this case, the
stability windows closes down, making clear, once again, the key role
played by the background pressure and energy density of the medium
which have to be different from zero to
avoid instabilities in the presence of six DoF.

\section{Stability in an Expanding Universe}
\label{FLRW}
Let now consider the case when the perturbation wavelength is
comparable with the curvature scale; in this case the background
pressure and energy density makes the Universe expand and we enter in
the realm of cosmology. To study perturbations we consider a spatially
flat FLRW perturbed solution in the conformal Newtonian gauge, namely
\be
ds^2=a^2 \, \eta_{\mu \nu}\, dx^\mu dx^\nu + 2 \, a^2 \left [ \Psi(t,\,\vec x) \, dt^2+
\Phi(t,\,\vec x) \, d\vec{x}^2 \right ] = a^2 \left( \eta_{\mu \nu} +
h_{\mu \nu} \right)dx^\mu dx^\nu\, ;
\label{ngm}
\ee
while for the St\"uckelberg fields in the scalar sector we have
\be
\varphi^0=\phi(t)+\pi_0(t,\vec{x}),\qquad \varphi^a=x^a+\partial_a
\pi_L(t,\vec{x}) \, .
\label{ngs}
\ee
The mass paramaters  $\lambda_a$ and $M_a$ in a FLRW background the are similarly
defined as in~(\ref{LUm}); it is convenient to introduce an overall factor
$a^4$~\footnote{Notice that such overall factor is not present in the
  definition given in~\cite{classus}.} in the
quadratic Lagrangian corresponding to the square root of minus the
determinant of the background metric; namely 
\be
L^{(2)}_{\text{mass}} =\frac{\plm^2 \, a^4} {4} \left[\lambda_0^2 \;  h_{00}^{(ug)}{}^2+2\,\lambda_1^2\;  h_{0i}^{(ug)\,2}
    %h_{0i}^{(ug)}   
    -2\,\lambda_4^2 \; h_{00}^{(ug)}\,  h_{ii}^{(ug)} +\lambda_3^2 
\;  h_{ii}^{(ug)}{}^2  - \lambda_2^2\;   h_{ij}^{(ug)\,2} %\,  h_{ij}^{(ug)}
\right] \, .
\ee
The expressions of $M_a$ are given in appendix~\ref{medium}. The
conservation of the background EMT is 
equivalent to the  equation of motion for $\bar \varphi^0 = \phi$
\be
\label{eqphi}
\rho'=-3\;{\cal H}\;(\rho+p)\quad\Rightarrow \quad
  \hat{M}_0 \,  \phi ''=  \left(\hat{M}_0+3 \, \hat{M}_4\;\right)\;\mathcal{H}\; \phi ' \, .
 \ee
The quadratic Lagrangian for scalar perturbation in the Fourier basis 
can be unambiguously written as (see appendix \ref{stabcomp})
\be
\begin{split}
&L_s^{\text{FRW}} = \frac{1}{2} \, {Q'}^t \cdot \pmb{K}\cdot Q' +
\dot{Q}^t\cdot \pmb{D} \cdot Q - \frac{1}{2} \, Q^t \cdot \pmb{M} \cdot Q \, ;\\[.2cm]
&  \pmb{K}^t = \pmb{K} \,, \qquad  \pmb{M}^t= \pmb{M} \, , \qquad  \pmb{D}^t = -  \pmb{D} \, ,
\end{split}
\label{quad}
\ee
where $Q^t=(\Psi, \, \Phi, \, k^2 \, \pi_l , \, k \, \pi_0) $ and $'$
denotes the derivative with respect of the conformal time.  In our case we
have that
\be
\pmb{K} = \left(
\begin{array}{cccc}
 0 & 0 & 0 & 0 \\
 0 & -12 \, a^2 \, \plm^2 & 0 & 0 \\
 0 & 0 & a^4 \;\frac{\left[\hat{M}_1+( p+ \rho )\right]}{k^2} & 0 \\
 0 & 0 & 0 & \frac{2 \, a^4 \, \hat{M}_0}{k^2 \, \phi'{}^2} \\
\end{array}
\right)  \, ;
\ee 
\be
\pmb{D} = \plm^2 \, \left(
\begin{array}{cccc}
 0 & 6 \, a^2 \, \mathcal{H} & 0 & -\frac{ a^4 \, M_0}{k \, \phi '} \\
 -6 \, a^2 \,  \mathcal{H} & 0 & 0 & \frac{3 \,  a^4 \, M_4}{k \, \phi '} \\
 0 & 0 & 0 & - a^4 \frac{\left(M_1-2 \, M_4\right)}{2 \, k \, \phi '} \\
 \frac{ a^4 \, M_0}{k \, \phi'} & -\frac{3 \,  a^4 \,  M_4}{k \, \phi '} &
   a^4 \, \frac{ \left(M_1-2 \, M_4\right)}{2 \, k \, \phi '} & 0 \\
\end{array}
\right) \, ;
\ee
We  denoted by ${\cal H} = a'/a$ the Hubble parameter in conformal
time. The mass matrix $\pmb{M}$ is not very illuminating. 
Differently from the analysis involving the St\"uckelberg fields in flat
space, this time the effect of the background on the FLRW metric is
consistently taken into account and
\be
{\cal H}^2 = \frac{ a^2 \,  \rho}{6 \, \plm^2} \, , \qquad {\cal
  H}'=- \frac{a^2 ( \rho + 3 \,  p)}{12 \, \plm^2} \; .
\ee
The field $\Psi$ has an
algebraic equation of motion and can be
integrated out getting; the resulting Lagrangian, after suitable
integration by parts can be again put in the form (\ref{quad}). 
As a final step, a second field can be integrated out. Indeed, defining
\be
q= \Phi- \frac{{\cal H}}{\phi'} \, \pi_0 \, ,
\ee
when the quadratic action is expressed in terms of $q$, $\pi_l$ and
$\pi_0$, it turns out that $\pi_0$ can be integrated out.
The action has the very
same general form (\ref{quad}) but with 2$\times$2 matrices:
\be\label{kin}
\begin{split}
& \pmb{K} =\frac{\plm^2}{f} \, \left(
\begin{array}{cc}
 12 \, a^2 \, \hat{M}_0 \left(3 \; a^2 \, \mathcal{M}_1 +4 \; \plm^2 \,
  k^2\right) & 12 \, a^4 \, \hat{M}_0
   \, \mathcal{M}_1 \\
 12 \, a^4 \, \hat{M}_0 \, \mathcal{M}_1 & 4 \;
   \rho \; a^4 \, \mathcal{M}_1 \\
\end{array}
\right) \, ; \\[.2cm]
& {\cal M}_1 = \hat{M}_1 + \rho+p  \, , \qquad f =4 \; \plm^2 \; k^2 \,
\rho -3 \, a^2 \, \mathcal{M}_1 \left(\rho -\hat{M}_0\right) \, ;
\end{split} 
\ee
and
\be
\pmb{D} = 12 \, a^2 \, k^2 \, \plm^4 \begin{pmatrix} 0 & - \frac{\left(\hat{M}_1-2
    \, \hat{M}_4\right) \mathcal{H}}{f} \\
\frac{\left(\hat{M}_1-2
    \, \hat{M}_4\right) \mathcal{H}}{f} & 0 \\ 
\end{pmatrix}  \, .
\ee
The expression of $\pmb{M}$ is not particularly illuminating and will not
be given here. What is  important is to find the condition for
which
the total energy is positive definite in the large $k$ limit, namely
when $k^2
\gg {\cal H}^2$,  to confirm the pure St\"uckelberg analysis in flat space. Proceeding as for the case of flat
space, in the  large $k$ limit, we find the  kinetic energy is positive if 
\be
M_0 >0 \qquad \text{and} \qquad 
   \hat{M}_1 + ( p+ \rho )  >0 \, .
\ee
Moreover on can easily see that such condition
together with 
\be
\hat{M}_0 < \rho \, 
\ee
enforce that the kinetic energy is positive definite for any $k$.
Turning to the  mass matrix in the energy, it is positive if
\be
M_2 >M_3 \qquad \text{and} \qquad  M_1 <0 \, .
\ee
Finally, looking at vector modes it is easy to see that~\cite{classus}
stability requires
\be
 \hat{M}_1 +  ( p+ \rho ) >0 \qquad \text{and} \qquad  M_2 >0 \, .
\ee
As a result, the stability conditions are the same of the ones derived
with the previous
St\"uckelberg analysis; in addition when $\hat{M}_0 < \rho $, ghost
instability are forbidden at any $k$. The only instability found is a
Jeans instability when, for  $k<k_{J} $, one of the eigenvalues of the
mass matrix of the total energy becomes negative. Such
gravitational instability is typical of any self-gravitating system. 

As a technical comment, we note that taking the limit ${\cal H} \to 0 $, via the
background equations, one is led again to $p=\rho=0$ and we are back
to well known problems discussed in the previous section. 

Let us consider a number of concrete examples.
\begin{itemize}
\item 
{\bf Perfect fluid}\\
This is the simplest case one can immagine. Flat space stability
actually coincide with thermodynamical stability~\cite{usthermo}. 
From the fluid Lagrangian of the form $U(,b,Y)$, we have that
$M_1=M_2=0$. The transverse spin 1 modes has a degenerate dispersion
relation related to the classical conservation of vorticity in a
perfect fluid and the dynamics of tensor modes is not
modified. Stability conditions are simpler and come only form
(\ref{energy}) with $M_1=0$ and are equivalently from the corresponding
expression in FLRW. Thus
\be
M_0 = \frac{\phi'{}^2 \, U_{YY}}{ 2 \, a^2 \, \plm^2 }>0, \qquad
\qquad p+\rho >0 \, ;
\ee
which can be easily satisfied. The only exception is when the null
energy condition is violated.
\item
{\bf Superfluid}\\
A superfluid is the simplest example of a medium with $M_1 \neq
0$. Being still a fluid, $M_2=0$. The Lagrangian has the general form
$U(b,Y,X)$, see appendix \ref{medium} and~\cite{classus} for the definition of the
relevant operators. Clearly the most stringent stability conditions
are the ones involving $M_1$ that for a superfluid is given, together
with the background value of
$X$ on FLRW, by
\be
M_1= \frac{2 \, \phi'{}^2 \, U_X}{\plm^2 \, a^2}, \qquad X =
-\frac{\phi'{}^2}{a^2} \,.
\ee
Moreover for superfluids ~\cite{classus}, see appendix~\ref{medium},  we have that
\be
p+\rho=\frac{\phi' \, U_Y}{a} -\frac{2 \, \phi'{}^2 \, U_X}{a^2} -
\frac{U_b}{a^3} \, ;
\ee
Then the stability condition involving $M_1$ reads simply
\be
U_X < 0,  \qquad \qquad 
\frac{\phi' \, U_Y}{a} - \frac{U_b}{a^3} >0 \, ;
\ee
remarkably there is a perfect cancellation of $U_X$ between $M_1$ and $\rho+p$
rendering such condition easy to satisfy. Notice that
$a^{-1} \, \phi' \, U_Y - a^{-3} \, U_b$ is just $p+\rho$ for the
normal component of the superfluid, see appendix \ref{medium}.
\item
{\bf Supersolid}\\
Take a supersolid (not the most general one) described by the
Lagrangian $U(b,Y,X,\tau_n)$ with $n=1,2,3$. The presence of
${\tau_n}$ turns on $M_2$, however the cancellation of $U_X$ in the
conditions involving $M_1$ still takes place and we have that
\be
U_X < 0,  \qquad \qquad \frac{\phi' \, U_Y}{a} - \frac{U_b}{a^3} -2
\left(\frac{U_{\tau_1}}{a^3}+ \frac{2 \, U_{\tau_2}}{a^4}+
      \frac{3 \, U_{\tau_1}}{a^3} \right) >0 \, ,
\ee
which again is not very restrictive. The condition is just   $\rho+p$ for
a finite temperature solid; thus, as for superfluids, the effect of
$M_1$ is to cancel the contribution from the superfluid component, see appendix \ref{medium}. 
\item
{\bf Most general supersolid}\\
Consider now the case of the moste general solid whose Lagrangian $U$
contains all ten operators rotational invariant operators, see
appendix \ref{medium}. While terms of the form $U_{y_m}$, $m=0,1,2,3$,
contribute to $M_1$,  there is no contribution in $p+\rho$ being of
the form $y_m \, U_{y_m}$ and $y_m=0$ at the background level. As a
result the stability conditions can be rather restrictive for the most
general supersolid.
\end{itemize}
The bottom line is that the stability conditions are easily satisfied
with possible exception of the most general supersolid where the
restriction on $M_1$ is non-trivial. It is also worth to stress the we
have studied the stability conditions in the worse case where the
medium itself is dominating the Universe and thus the scale of $\rho$
and $p$ are the same of $\hat{M}_a$. For instance, when radiation
is dominant we expect that very similar stability conditions should
hold with $p$ and $\rho$ associated with the dominant component
(ultra-relativistic particles) and  $\hat{M}_a \ll p+\rho$, making
stability  not an issue.
 
\section{Massive Gravity Phases}
\label{phases}
Here we briefly describe what happens once the self-gravitating nature of the medium has been taken into
account. 
The special case of  a de Sitter background, where $p+\rho=0$,
which incidentally is naturally the one selected by the Lorentz
invariant (LI)  quadratic Lagrangians, will be analysed in a forthcoming paper;  here
we just report, for completeness, the main 
results.
  
\begin{itemize}
\item The most
interesting and new case is when basically no tuning on $U$ is required and
this also applies to the resulting masses.\footnote{In particular we have to avoid the zeros in the kinetic matrix \eq{kin}, ${\cal M}_1,\,M_0\neq0$.} The vacuum state breaks
Lorentz invariance  which in turn allows the existence of
non-trivial pressure and energy density background; stability can be
achieved when $p+\rho \neq 0$. The full six DoF of the theory can be stable at quadratic order . 

\vspace{0.3cm}

There are two critical points in the parameter space where the kinetic
matrix is degenerate and the above
analysis has to be repeated (we always require $p,\,\rho\neq 0$) 

\item 
When $\hat M_0=\plm^2\;\l_0^2-\rho=0$ we have at least one DoF less in the scalar sector, for a total of 5 DoF.
The conservation of the background
EMT is equivalent to \eq{eqphi}  and 
for  $\hat M_0=0$ (with ${\cal
  H}\neq0$) it implies that  also $\hat
M_4=\plm^2\;\lambda_4^2-p/2=0$. 
No constraint on $\hat M_4$ exists on a Minkowski background where ${\cal H}=0$.
 
\begin{itemize}
\item 
In the Lorentz breaking (LB)  case we have only one propagating scalar whose kinetic term is proportional to
$k^2\;{\cal H}^2\;(1+w)$ and mass to $(M_2-M_3)$ (in the large $k$
limit). Thus when $w=-1$ the kinetic term for such a mode vanishes.

\item
For a LI quadratic Lagrangian and with  $\rho+p=0$, the requirement of  $M_4=0$, for the existence of an FLRW background, gives
\be
\label{LI0} 
\begin{cases} 
M_{0}=A+B-\frac{\rho}{2}=0  \\
M_{3,4}=B+\frac{\rho}{2}=0  \\
\end{cases}
 \Rightarrow M_{1,2}=-A+\rho=0 \, .
\ee
As a result, a LI self gravitating medium in such a phase has all masses
$M_i$ zero and the parameters $A$ and $B$ are 
 fine tuned according with: $\plm^2 \, A=\rho$, $\plm^2 \, B=-\rho/2$.
 and no scalar and vector propagates at the quadratic order. Only two
 massless  tensor modes are present.
In Minkowski space, keeping  $M_4\neq 0$ also $M_1\neq 0$ and then
tensors and transverse vectors propagate and we get 5 DoF.  In the contest of the decoupling limit of
Pauli-Fierz massive gravity, the vanishing of the kinetic term in the
scalar sector is dealt by changing the St\"uckelberg content by adding
an additional scalar~\cite{ArkaniHamed:2002sp,Hinterbichler:2011tt}. 
\end{itemize}

\item When $\hat M_1+\rho+p=\plm^2\;\l_1^2+\rho=0$, only a single scalar and two tensor modes
  propagate for a total of $\leq 3$ DoF; such a phase can be implemented in a non-perturbative way and
can be protected by a symmetry \cite{Rubakov:2008nh,Dubovsky:2004sg}. 
\begin{itemize}
\item In the LB theory kinetic term for the only propagating scalar mode is given by
\be
\label{kinm1}
K=\frac{1}{2\;{\cal H}^2}\left[
a^4\; \hat{M}_0+\frac{(\hat{M}_4\;a^2+3\;(1+w)\;{\cal H}^2)^2}{\hat{M}_2-\hat{M}_3}
\right] \, ,
\ee
that is positive defined for $M_0>0,\;M_2>M_3$.
 
\item
In the LI case being $\l_1^2=-A$ implies $A=\rho$ and, being  
$p+\rho=0$, we have
\be
M_{1,2}=0,\qquad \hat{M}_{0,3,4}=B+\frac{\rho}{2} \, .
\ee
From \eq{kinm1} we get $K=0$ and then  no scalar mode propagates.
Likewise in GR, only massless tensor mode are present.
\end{itemize}

\end{itemize}

Let us comment on the Higuchi
bound~\cite{Higuchi:1986py,Higuchi:1986wu}. Such a bound on the
Pauli-Fierz mass in dS spacetime is derived when the massive spin 2
action gives no contribution to the
background~\cite{Deser:2001wx,Grisa:2009yy}, namely no linear term is
present. In the dRGT
massive gravity model, a dS  background exists when a non-flat time-dependent
reference metric~\cite{Fasiello:2012rw} (basically the same of the one
found in the contest of 
bigravity~\cite{Comelli:2011zm,Comelli:2012db}) is considered; in this
case the Eguchi bound is recovered. In our approach, by
definition, the medium is self-gravitating and the above considerations
do not apply.

%%%%%%%%%%%%%%%%%%%%%%%%%%%%%%%%%
%%%%%%%%%%%%%%%%%%%%%%%%%%%%%%%%%%
\section{Conclusions}
\label{conc}
We  have analysed   the low energy phonon-like excitations
of generic media  (fluids, superfluids, solids and supersolids) on Minkowski space
showing that typically no instability is present. 
 It turns out that when dynamical gravity is added to the game, in the unitary gauge,
self-gravitating media are equivalent to massive gravity and six
degrees of freedom are present. 
On the other hand, massive gravity
theories with six DoF on Minkowski are plagued by ghost instabilities and a great
effort has been devoted trying to find a non-perturbative way to
project out the unwanted (ghost) sixth mode. 
We reconcile the two
apparent contradictory facts by taking into account the pivotal role played by 
the pressure and energy density of the medium.
 The point is that flat space is
not a consistent background for a self-gravitating medium and for
massive gravity, unless the background pressure and energy density is
set to zero. 
In such case the generic stability conditions for a medium are
badly violated. This is not surprising, when $p=\rho=0$ the
medium picture breaks down simply because there is no medium.
To consistently study the dynamics of a self-gravitating media in  
presence of dynamical gravity one has to change background taking a 
FLRW Universe; the stability analysis around such background again
shows that generically no ghost instability is present and actually
for modes with wavelength much smaller   of the curvature scale
${\cal H}^{-1}$ the flat space results with gravity switched off are
recovered.
Interestingly, the fluid picture is incompatible with
the requirement of Lorentz invariance of the medium energy-momentum
tensor and leads again to $p=\rho=0$ and inevitably the Lagrangian has
to be  tuned to get less than six DoF.
 As a result, the
spontaneous breaking of Lorentz invariance is instrumental to open up
the stability window for massive gravity/self-gravitating medium with
six DoF. Such a breaking is an emergent low-energy phenomenon triggered by 
the non-vanishing homogenous background
value of the St\"uckelberg scalar fields due to the presence of a medium with non zero energy density and pressure.
One of the main theoretical
and phenomenological features of such a models is that the Einstein-Hilbert
part of the action is untouched keeping the dynamics of the spin two
graviton very close to GR, except for a massive deformed dispersion
relation
\be
\o=\sqrt{k^2+m^2}\simeq k+\frac{m^2}{2\;k}+...\qquad k^2\gg m^2 \, .
\ee
For  a  typical frequency   of the order some Hertz,
  observations~\cite{TheLIGOScientific:2017qsa,Monitor:2017mdv}
  require that the graviton velocity is very close to $c=1$:
\be
\Delta v \sim\frac{m^2}{k^2}\leq 10^{-15}\quad \Rightarrow \quad m<10^{-22}\;eV
\, .
\ee
Such a bound on the mass  is not as stringent as requiring that 
the mass scale is of the order  of  the present Hubble parameter $\sim
10^{-33}\;eV$ if massive gravity is relevant for dark energy.

%\newpage

%%%%%%%%%%%%%%%%%%%
  %%%%%%%%%%%%%%%%%%%
 \begin{appendix}

\section{Self-Gravitating  Media}
\label{medium}
The operators entering the Lagrangian describing the low energy
dynamics of a general isotropic medium minimally coupled with gravity can be given in terms of matrix~\cite{ussgf,classus}
\be
C^{AB}= g^{\mu \nu} \, \de_\mu \varphi^A \de_\nu \varphi^B,\qquad A,B=0,1,2,3
\ee
by
\be
\begin{split}
&X= C^{00} \, , \qquad \tau_a =\text{Tr} \left(\pmb{B}^a \right) \; \; 
a=1,2,3 \, , \qquad y_n  =\text{Tr} \left(\pmb{B}^n  \cdot \pmb{Z}
\right) \; \;  n=0,1,2,3 \, ;\\[.2cm]
& \left(\pmb{B} \right)^{ab}= C^{ab} \equiv B^{ab} \, , \qquad
 \left(\pmb{Z} \right)^{ab}= C^{0b}
\, C^{0b} \, .
\end{split} 
\ee
For a perfect fluid $U$ can be taken as  a function of only two operators $b$ and $Y$ defined by 
\be
b=\sqrt{\det \left(B^{ac}
  \right)} ,\, ;  a,c=1,2,3  \qquad \qquad\, u^\mu=\frac{\epsilon^{\mu\alpha \beta
  \gamma}\;\partial_\alpha  \varphi^1\;\partial_\beta
\varphi^2\;\partial_\gamma\varphi^3}{b\;\sqrt{g} } \, , \qquad \qquad  Y =
u^\mu \, \nabla \varphi^0 \,  .
\ee
The 4-vector $u^\mu$ is the fluid's velocity and $u^\mu \de_\mu
\varphi^a =0$; thus $\varphi^a$ can be
interpreted as the spatial Lagrangian (comoving) coordinates of the fluid,
while $\varphi^0$ represents the fluid's clock's. Together with $b$,
and $Y$, $X$, $\{\tau_n \}$ and $\{y_m \}$ form a set of ten
independent rotational invariant operators, thus the
Lagrangian of a general medium will have the form $U(b,Y,X,\tau_a,y_n)$.
The EMT tensor for a general medium as the form
\be
 \begin{split} 
&T_{\mu \nu} = \rho \, u_{\mu} \, u_{\nu} + q_\mu \, u_\nu + q_\nu \,
u_\mu+ {\cal P}_{\mu \nu} \, ;\\
& \rho = T^{\mu \nu}\, u_\mu \,u_\nu \, , \qquad q_\mu =- h_{\mu \alpha}\,
T^{\alpha \beta} \,u_\beta \, , \qquad {\cal P}_{\mu \nu} = h_{\mu
 \alpha} \,h_{\nu \beta}\, T^{\alpha \beta} \, ;
\end{split}
\label{sgmemt}
\ee
where $h_{\m\n}=u_\m\,u_\n+g_{\m\n}$ is the projector orthogonal to
$u^\mu$ and 
\be
q_\mu = 2\;Y \left[   \, \sum_{m=0}^3 U_{y_m}
  \left(\boldsymbol{B}^m \right)^{ab} \, \nabla_\mu \varphi^b \, C^{a0}
-  \, U_X \, \xi_\mu \right] \, .
\ee
has the form of the heat flow vector while splitting
\be\label{Tmn2}
{\cal P}_{\mu \nu}= h_{\mu \nu} \, p + {\cal P}_{\mu
  \nu}^{\text{tl}} \, , \qquad  p = \frac{ {\cal P}_{\mu \nu}
    h^{\mu \nu}}{3} \, , \qquad  {\cal P}_{\mu
  \nu}^{\text{tl}} h^{\mu \nu} =0 \ ;
\ee
one has
\ba
\label{rhop}
 && \rho= -U+Y\;U_Y-2\;Y^2\; U_X \, ;\\  
 && p= U-b\;U_b-\frac{2}{3}\;\sum_{n=1}^3\;n\;\tau_n\;U_{\tau_n}
-\frac{2}{3}\,\sum_{n=0}^3\;n\;y_n\;U_{y_n}-\frac{2}{3}\;U_X\;(Y^2+X)
\, .  
 \ea
In the case of a perfect fluid we simply have that $ \rho= -U+Y \,U_Y$
and $p=U-b\;U_b$.

\be
 \rho= -U+Y \,U_Y , \qquad p=U-b\;U_b \, .
\label{prho}
\ee
For explicit expressions for the $\lambda_a$ are the following
\be
\begin{split}
&\lambda_0^2=\frac{-a^4 \, U+a^2 \,\phi'{}^2 \left(U_{YY}-4
   \, U_X\right)+a \, \phi'  \, U_Y -4 \, a \, \phi'{}^3 \,
   U_{YX} +4 \, \phi'{}^4  \, U_{XX} }{2 a^4 \plm^2} \, ;\\
&\lambda_1^2= \plm^{-2} \left( U- a^{-3} \, U_b +2 \, a^{-2} \, \phi'{}^2
  \, U_X - \sum_{m=1}^3 2 \, m \, a^{-2m} \, U_{\tau_m} -2 \,
  \phi'{}^2 \sum_{n=0}^3 a^{-4-2n} \, U_{y_n} \right) \, ;\\
& \lambda_2^2 = \plm^{-2} \left[U - a^{-3} \, U_b- 4 \left(a^{-2} \,
    U_{\tau_1} + 3 \, a^{-4} \, U_{\tau_2} + 6 \, a^{-6} \, U_{\tau_3}
  \right) \right] \, ;\\
&\lambda_3^2= \frac{1}{2 \, \plm^2} \left[ U - a^{-3} \, U_b + a^{-6} \,
U_{bb} - 4 \sum_{m=1}^3 n \, a^{-2n} \, U_{\tau_n}  + 4 \, a^{-3}
\sum_{m=1}^3 n \, a^{-2n} \, U_{b \tau_n} +\frac{36 \,  U_{\tau
    _3^2}}{a^{12}}+\frac{48 \, U_{\tau _2 \tau
   _3}}{a^{10}} \right.\\
&\qquad \left. +\frac{16 \, U_{\tau _2^2}}{a^8}+\frac{24 \, U_{\tau _1
   \tau _3}}{a^8}+\frac{16 \, U_{\tau _1 \tau _2}}{a^6}+\frac{4
   U_{\tau _1^2}}{a^4}\right] \, ;\\
&\lambda_4^2= \frac{1}{2 \, \plm^2} \left[ U - a^{-3} \, U_b -
    \frac{\phi'}{a} \, U_Y +2 \, 
    \frac{\phi'{}^2}{a^2} \, U_X +
    \frac{\phi'}{a^4} \, U_{bY} -
2\,     \frac{\phi'^2}{a^5} \, U_{bX}  -2 \sum_{n=1}^3 n \, a^{-2n}\,
U_{\tau_n} \right.\\
&\qquad \left.  -4 \, \frac{\phi'{}^2}{a^2} \sum_{n=1}^3 n \, a^{-2n} \,
U_{X\tau_n}+ \frac{\phi'}{a} \sum_{n=1}^3 2 \, n \, a^{-2n} \,
    U_{Y\tau_n} \, . \right]\label{LLI}
\end{split}
\ee
The mass parameters $\{M_a\}$ are defined separating the contribution
from $\{\lambda_a \}$ the part that is related to the background
pressure and energy density, this is done by~(\ref{MM}) which leads
to
\be
\begin{split}\label{MMI}
& M_0= \frac{\phi'{}^2}{2 \, a^4 \, \plm^2} \left[a^2 \left(U_{YY}-2
    \, U_X\right)-4\,  a \, \phi' \, U_{YX} +4\, \phi'{}^2 \, 
   U_{XX} \right] \, ;\\
&M_1 = \frac{2 \, \phi'{}^2}{\plm^2} \left[ a^{-2} \, U_X
  +\sum_{n=0}^3 a^{-4-2n} \, U_{y_n}\right] \, ;\\
&M_2 = -\frac{2}{\plm^2} \sum_{m=1}^3 n^2 \, U_{\tau_n} \, ;\\
&M_3 = \frac{1}{\plm^2} \left( 
     2  \,\sum_{m,n=1}^3 m\, n\,
   a^{-2 m-2 n} \;U_{\tau _m \tau
   _n}+2  \; \sum_n\;n\;  a^{-3-2 n}\;
   U_{b \tau _n}- \;
    \sum_{n,m=1}^3 
   \; a^{-2 n}\; U_{\tau _n}+
   \frac{1}{2\,a^6}\;U_{b^2}  \right) \, ;\\
&M_4=  \frac{1}{\plm^2} \left[\frac{\phi'}{2} \;\left(2\;  \sum_{m,n=1}^3 
   a^{1-2 n}\; U_{Y \tau
   _n}-\frac{U_Y}{a}+  \frac{U_{bY}}{a^4}\right)+
  \phi'^2 \;\left(-2\;  \sum_{n=1}^3\,  a^{-2-2 n}
   \;U_{X \tau _n}+  \frac{U_{X}}{a^2}- \frac{ U_{b X}}{a^5}\,
  \right) \right]\, ;
\end{split}
\ee
In Minkowski space the mass parameters are obtained form the above
expression by setting $a=\phi'=1$.

\section{Unitary vs Newtonian gauge in FRW space}
\label{stabcomp}
The quadratic action for scalar perturbations in a Fourier basis 
can be unambiguously written in matrix notation as 
\be
{\cal L}_2^{ } =
% \int\left.\sqrt{-g}\;M_{pl}^2\;R+ \sqrt{-g}\;U \right |_{{\cal O}(2)}
% \equiv 
\frac{1}{2} \, {Q'}^t \cdot {\pmb K} \cdot Q' +
{Q'}^t \cdot {\pmb D} \cdot Q - \frac{1}{2} \, Q^t \cdot {\pmb M} \cdot Q \, ;
\label{quadd}
\ee
the relevant metric and scalar field perturbations are collected in
$Q$,  a vector with 4 components.  
Then we have the 4$\times$4 matrices:\footnote{If the matrix $D$
  in \eq{quadd} is not antisymmetric,  it is possible by an
  integration  by parts to kill its symmetric part; as consequence, in
  \eq{quadd}  one has to perform the following replacement: $\pmb{D}
  \to\frac{\pmb{D}-\pmb{D}^t}{2}$ 
and $\pmb{M} \to \pmb{M}+\frac{\pmb{D}'+{\pmb{D}'}^t}{2}$.}
The matrix ${\pmb K}$ is  the kinetic matrix and its symmetric by
construction as the mass matrix $\pmb{M}$.
If a non dynamical field is present, it can be integrated out and the
resulting will be still of the form (\ref{quadd}) with a new $Q$ and
$\pmb{K}$, $\pmb{D}$ and $\pmb{M}$ of lower dimension.
In order to study the stability of  system described by
(\ref{quadd}) we simply impose the positivity of the Hamiltonian.
 The conjugate momenta are
\be
\begin{split}
&
\Pi=\frac{\partial {\cal L}_2}{\partial {Q'}^{t} }=\pmb{K} \cdot Q'
+\pmb{D} \cdot Q \,.
\end{split}
\ee
When ${\pmb K}$ is non-degenerate we are sure that no constraint is
present. The total energy $E$ , basically the Hamiltonian, expressed as a
quadratic form in $Q'$ and $Q$ is 
\be
E =  \;{Q'}^{t} \cdot \left(\pmb{K}\cdot Q' +\pmb{D}\cdot Q\right) -{\cal L}_2=
\frac{1}{2} \left( {Q'}^t \cdot \pmb{K} \cdot Q' + Q^t \cdot \pmb{M}
  \cdot Q \right)  \, .
\ee
Equivalently, the Hamiltonian in terms of $\Pi$ and $Q$ can be written as
\be
H =   \ha \left[ (\Pi^{t}-  Q^t \cdot \pmb{D}^t) \cdot \pmb {K}^{-1} \cdot(\Pi-
  \pmb{D} \cdot  Q)+Q^t \cdot\pmb{M} \cdot Q \right]\, .
\label{ham}
\ee
It is then  clear that the positivity of the
Hamiltonian or the energy is equivalent to the  positivity of the kinetic matrix $\pmb{K}$ and
mass matrix $\pmb{M}$,  independently from $\pmb{D}$.
% When we are in presence of only second class constraints we can
% integrate out the auxiliary fields and  
% start again from the previous discussion with new reduced matrices.

%%%%%%%%%%%%%%%%%%%%%%%%%%%%%%%%%%%
%%%%%%%%%%%%%%%%%%%%%%%%%%%%%%%%%%%
\subsection{Unitary Gauge}
\label{SU}
In the unitary gauge
   all  scalar perturbations in the metric according with 
\be
\label{hU} 
ds^2 =a^2\; \left(\eta_{\mu \nu}+ h^{(U)}_{\mu \nu} \right) dx^\mu dx^\nu ,\,
\qquad h_{00}^{(U)} = 2  \, \Psi ,\, \quad h_{0i}^{(U)} = \de_i v ,\,   \quad h_{ij} ^{(U)}=
2 \,\Phi\; \delta_{ij} + \de_i \de_j \sigma \, ;
\ee
while the scalars fields are frozen in their background value
\be
\label{piU}
 \varphi^0 = \phi(t)\,, \qquad \varphi^a =x^a\;\;\;a=1,2,3 \, .
\ee
The relevant fields are
$Q^{(U)}=(\Psi,\;\Phi,\;v,\;\sigma)$ and  from the expansion of (\ref{factg}) we get
\be
\begin{split}\label{KU}
& {\pmb K}^{(U)} = \plm^2 \left(
\begin{array}{cccc}
 0 & 0 & 0 & 0 \\
 0 & -12  & 0 & 2  \, k^2  \\
 0 & 0 & 0 & 0 \\
 0 & 2  \, k^2  & 0 & 0 \\
\end{array}
\right) \, , \qquad {\pmb D}^{(U)} = \plm^2 \left(
\begin{array}{cccc}
 0 & 0 & 0 & 0 \\
 0 & 0 & -2 \, k^2 & 0 \\
 0 & 2 \, k^2 & 0 & 0 \\
 0 & 0 & 0 & 0 \\
\end{array}
\right) \, , \\[.2cm]
& {\pmb M}^{(U)} = \plm^2 \left(
\begin{array}{cccc}
 -2 \lambda _0 & 2 \left(2  \,  k^2+3  \, \lambda _4\right) & 0 & -k^2
   \lambda _4 \\
 2 \left(2  \, k^2+3  \, \lambda _4\right) & -2 \left(2  \, k^2-3  \, \lambda _2+9 \, 
   \lambda _3\right) & 0 & -k^2 \left(\lambda _2-3  \, \lambda
   _3\right) \\
 0 & 0 & -k^2 \, \lambda_1 & 0 \\
 -k^2  \,  \lambda _4 & -k^2 \left(\lambda _2-3  \, \lambda _3\right) & 0 &
   \frac{1}{2} k^4 \left(\lambda _2-\lambda _3\right) \\
\end{array}
\right) \, .
\end{split}
\ee
Notice the in the unitary gauge in the quadratic Lagrangian nor $p$
neither $\rho$ si present, thus imposing the background equation of
motions $p=\rho=0$ is immaterial. From the form of $K$ is clear that
both $\Psi$ and $v$ are auxiliary field and can be integrated out when
$\lambda_0 \neq 0$ ans $\lambda_1 \neq 0$. After integrating out those
fields we arrive at an effective quadratic Lagrangian still of the
form of (\ref{quad}) with $Q=(\Phi, \sigma)$ with
\be
\pmb{K}_2^U= \plm^2 \left(
\begin{array}{cc}
 -4 \left(4  \, k^2+3  \, \lambda _1\right)/\lambda _1 & 2  \, k^2 \\
 2 \,  k^2 & 0 \\
\end{array}
\right) \, , \qquad \pmb{D}_2^U= 0 \, .
\ee
The expression of the mass matrix is not relevant for us. Clearly,
there are two propagating DoF and one is a ghost.

\subsection{Newton  Gauge}
\label{SN}
In the Newtonian gauge the scalar  perturbations are both in the
metric and in the scalar fields, namely
 \be
\label{hN}
 h_{\m\n}^{(N)}=2\;\Psi\;\delta_\m^0\;\delta_\n^0+2\;
 \Phi\;\delta_\m^i\;\delta^i_\n \, ;
 \ee 
and 
 \be
\label{GN}
\varphi^0=\phi(t)+\pi_0,\;\varphi^a=x^a+\partial_a\;\pi_l \, ;
  \ee
In the Newtonian gauge   $Q^{(N)}=(\Psi, \, \Phi, k^2 \, \pi_l , k \,  \pi_0) $ and
\bea\label{KN}
&&{\pmb K}^{(N)} = \left(
\begin{array}{cccc}
 0 & 0 & 0 & 0 \\
 0 & -12  \, \plm^2 & 0 & 0 \\
 0 & 0 & 2  \,  \plm^2 \, \lambda_0-\rho  & 0 \\
 0 & 0 & 0 & k^2  \left(\plm^2 \, \lambda_1 +\rho \right) \, ;\\
\end{array}
\right)
\\&&
{\pmb D}^{(N)} = \left(
\begin{array}{cccc}
 0 & 0 & 0 & \frac{1}{2} \left(\rho -2  \, \plm^2  \, \lambda _0\right) \\
 0 & 0 & 0 & 3 \,  \plm^2  \, \lambda _4-\frac{3  \, p}{2} \\
 0 & 0 & 0 & -\frac{1}{2} k^2 \plm^2 \left(\lambda _1-2 \lambda
   _4\right) \\
 \plm^2  \, \lambda _0-\frac{\rho }{2} & \frac{3}{2} \left(p-2  \, \plm^2
   \lambda _4\right) & \frac{1}{2} k^2  \, \plm^2  \, \left(\lambda _1-2 \, 
   \lambda _4\right) & 0 \\
\end{array}
\right) \, .\\ 
&&
\frac{\pmb{M}^{(N)}}{\plm^2} =  \left(
\begin{array}{cccc}
 -2 \, \lambda _0 & 2 \, \left(2 \, k^2+3 \, \lambda _4\right) & k^2
   \left(2 \, \lambda _4 +\frac{\rho}{\plm^2} \right) & 0 \\
 2 \left(2 \,  k^2+3 \, \lambda _4\right) & 2  \left(3 \, \lambda_2-9 \,  \lambda_3 -2\,
 k^2\right) & k^2\left[2
\left(\lambda_2-3 \, \lambda _3\right)+\frac{p}{ \plm^2}\right] & 0 \\
 k^2 \left(2 \, \lambda _4 + \, \frac{\rho}{\plm^{2}} \right) & k^2 \left[2
   \left(\lambda _2-3 \, \lambda _3\right) + \frac{p}{\plm^2}\right] & k^4
   \left[2 \left(\lambda _2-\lambda _3\right) -\frac{p}{\plm^2}\right] & 0 \\
 0 & 0 & 0 & k^2 \left(\frac{p}{\plm^2}-\lambda _1\right) \\
\end{array}
\right) . \nb
\ea
Notice that this time the
quadratic Lagrangian depends on $p$ and $\rho$ if the background
equations are not imposed. Now $\pmb{K}$ is rank three and there are three
DoF in sharp contrast with the unitary gauge computation. The only
way to recover the same result if to impose $p=\rho=0$; then also
$\pi_0$ can be integrated out and the we end up with two DoF and one is
a ghost.

\end{appendix}

  \bibliographystyle{JHEP}  
  
%\bibliography{6thbiblio.bib}
\bibliography{6thbiblio.bib}

\end{document}